\documentclass[12pt]{article}
\usepackage{epsf}
\title{ {\bf
Lepton flavor violating $Z\rightarrow l_1^+ l_2^-$ decay in the
two Higgs Doublet model with the inclusion of non-universal extra
dimensions.}}
\author{\vspace{1cm}\\
        {\bf E. O. Iltan}
        \thanks{E-mail address:
        eiltan@heraklit.physics.metu.edu.tr}
 \\
        Physics Department, Middle East Technical University \\
        Ankara, Turkey\\}

\date{}

\begin{document}
\setlength{\baselineskip}{24pt}
\maketitle
\setlength{\baselineskip}{7mm}
\begin{abstract}
We predict the branching ratios of $Z\rightarrow e^{\pm}
\mu^{\pm}$, $Z\rightarrow e^{\pm} \tau^{\pm}$ and $Z\rightarrow
\mu^{\pm} \tau^{\pm}$ decays in the model III version of the two
Higgs doublet model, with the inclusion of one and two spatial
non-universal extra dimensions. We observe that the the branching
ratios are not sensitive to a single extra dimension, however,
this sensitivity is considerably large for two extra dimensions
\end{abstract}
\thispagestyle{empty}
\newpage
\setcounter{page}{1}
\section{Introduction}
The lepton flavor violating (LFV) interactions are interesting in
the sense that they are sensitive the physics beyond the standard
model (SM) and they ensure considerable information about the
restrictions of the free parameters, appearing in the new models,
with the help of the possible accurate measurements. Among LFV
interactions, the Z decays with different lepton flavor outputs,
such as $Z\rightarrow e \mu$, $Z\rightarrow e \tau$ and
$Z\rightarrow \mu \tau$, are rich enough to study and there is an
extensive work related to these decays in the literature
\cite{Riemann}-\cite{Perez}. The Giga-Z option of the Tesla
project which aims to increase the production of Z bosons at
resonance \cite{Hawkings} stimulates one to make theoretical works
on such Z decays.

In the framework of the SM the lepton flavor is conserved and, for
the flavor violation in the lepton sector, there is a need to
extend the SM. One of the candidate model is so called $\nu$SM,
which is constructed by taking neutrinos massive and permitting
the lepton mixing mechanism \cite{Pontecorvo}. In this model, the
theoretical predictions for the branching ratios (BRs) of the LFV
Z decays are extremely small in the case of internal light
neutrinos \cite{Riemann,Illana}
\begin{eqnarray}
BR(Z\rightarrow e^{\pm} \mu^{\pm})\sim BR(Z\rightarrow e^{\pm}
\tau^{\pm})
&\sim& 10^{-54}  \nonumber \, , \\
BR(Z\rightarrow \mu^{\pm} \tau^{\pm}) &<& 4\times 10^{-60} .
\label{Theo1}
\end{eqnarray}
They are far from the experimental limits obtained at LEP 1
\cite{PartData}:
\begin{eqnarray}
BR(Z\rightarrow e^{\pm} \mu^{\pm}) &<& 1.7\times 10^{-6} \,\,\,
\cite{Opal}
\nonumber \, , \\
BR(Z\rightarrow e^{\pm} \tau^{\pm}) &<& 9.8\times 10^{-6}\,\,\,
\cite{Opal,L3} \nonumber \, , \\
BR(Z\rightarrow \mu^{\pm} \tau^{\pm}) &<& 1.2\times 10^{-5} \,\,\,
\cite{Opal,Delphi} \label{Expr1}
\end{eqnarray}
and from the improved ones at Giga-Z \cite{Wilson}:
\begin{eqnarray}
BR(Z\rightarrow e^{\pm} \mu^{\pm}) &<& 2\times 10^{-9}  \nonumber \, , \\
BR(Z\rightarrow e^{\pm} \tau^{\pm}) &<& f\times 6.5\times 10^{-8}
\nonumber \, , \\
BR(Z\rightarrow \mu^{\pm} \tau^{\pm}) &<& f\times 2.2\times
10^{-8} \label{Expr2}
\end{eqnarray}
with $f=0.2-1.0$. Notice that these numbers are obtained for the
decays $Z\rightarrow \bar{l}_1 l_2+ \bar{l}_2 l_1$, namely
\begin{eqnarray}
BR(Z\rightarrow l_1^{\pm} l_2^{\pm})=\frac{\Gamma (Z\rightarrow
\bar{l}_1 l_2+ \bar{l}_2 l_1)}{\Gamma_Z} \, .
\end{eqnarray}
To enhance the BRs of the corresponding LFV Z decays some other
scenarios have been studied. The possible scenarios are the
extension of $\nu$SM with one heavy ordinary Dirac neutrino
\cite{Illana}, the extension of $\nu$SM with two heavy
right-handed singlet Majorana neutrinos \cite{Illana}, the Zee
model \cite{Ghosal}, the model III version of the two Higgs
doublet model (2HDM), which is the minimal extension of the SM
\cite{EiltZl1l2}, the supersymmetric models \cite{Masip, Cao},
top-color assisted technicolor model \cite{Yue}.

The present work is devoted to predictions of the BRs of
$Z\rightarrow e^{\pm} \mu^{\pm}$, $Z\rightarrow e^{\pm}
\tau^{\pm}$ and $Z\rightarrow \mu^{\pm} \tau^{\pm}$ decays in the
model III version of the 2HDM, with the inclusion of one and two
spatial extra dimensions. Our motivation is to check whether there
is an enhancement in the BRs of these decays due to the extra
dimensions. The possible existence of new dimensions reach great
interest recently and there is a large amount of work done in the
literature \cite{Dvali}-\cite{Yamaguchi}. The idea of extra
dimensions was originated from the study of Kaluza-Klein
\cite{Kaluza} which was related to the unification of
electromagnetism and the gravity and the motivation increased with
the study on the string theory which was formulated in a
space-time of more than four dimensions. Since the extra
dimensions are hidden to the experiments at present (for example
see \cite{Long}), the most favorable description is that these new
dimensions are compactified to the surfaces with small radii,
which is a typical size of corresponding extra dimension. This
leads to appear new particles, namely Kaluza-Klein (KK) modes of
the particles in the theory. In the case that all the fields feel
the extra dimensions, so called universal extra dimensions (UED),
the extra dimensional momentum, and therefore, the KK number at
each vertex is conserved. The compactification size $R$ has been
predicted as large as few hundereds of GeV \cite{ArkaniHamed,
Arkani, Antoniadis1,Antoniadis3}, in the range $200-500\, GeV$,
using electroweak precision measurements \cite{Appelquist}, the \(
B-\bar{B} \) -mixing \cite{Papavassiliou,Chakraverty} and the
flavor changing process $b \to s \gamma$ \cite{Agashe} for  a
single UED.

The assumption that the extra dimensions are at the order of
submilimeter distance, for two extra dimensions, the hierarchy
problem in the fundamental scales could be solved and the true
scale of quantum gravity would be no more the Planck scale but in
the order of electroweak (EW) scale \cite{Dvali,ArkaniHamed}. In
this case, the gravity is spreading over all the volume including
the extra dimensions, however, the matter fields are restricted in
four dimensions, called four dimensional (4D) brane, or in 4D
surface which has a non-zero thickness in the new dimensions,
called fat brane (see for example \cite{Rujula}). This type of
extra dimensions, accessible to some fields but not all in the
theory, are called non-universal extra dimensions (NUED). Contrary
to the UED, in the NUED, the KK number at each vertex is not
conserved and tree level interaction of KK modes with the ordinary
particles can exist. The study in \cite{Dienes} is devoted to the
appearence of a very light left handed neutrino in the NUED where
only the right handed neutrino is accessible to the extra
dimension. In the another work \cite{Agulia},  the effect of brane
kinetic terms for bulk scalars, fermions and gauge fields in
higher dimensional theories, have been studied. In \cite{iltanEDM}
the electric dipole moments of fermions and some LFV  decays have
been analyzed in the framework of NUED.

Here, we predict the BRs of the LFV Z decays in the model III with
the assumption that the extra dimensions are felt by the new Higgs
doublet and the gauge sector of the theory. The Z decays under
consideration are induced by the internal neutral Higgs bosons
$h^0$ and $A^0$ and their KK modes carry  all the information
about the new dimensions, after the compactification of the single
(double) extra dimension on the orbifold $S^1/Z_2$ ( ($S^1\times
S^1)/Z_2$. Here, the KK number in the vertices is not conserved,
in contrast to the UED case. The non-zero KK modes of neutral
Higgs fields $H$ have masses $\sqrt{m_{H}^2+m_n^2}$
($\sqrt{m_{H}^2+m_n^2+m_r^2}$) with $m_{k}=k/R$ in one (two) extra
dimension. We observe that the BRs of the processes we study
enhance almost two orders larger compared to the ones without the
extra dimensions, in the case of two NUED, since the neutral Higgs
KK modes are considerably crowded.

The paper is organized as follows: In Section 2, we present the
effective vertex and the BRs of LFV Z decays in the model III
version of the 2HDM with the inclusion of NUED. Section 3 is
devoted to discussion and our conclusions. In appendix section, we
give the explicit expressions of the form factors appearing in the
effective vertex.
\section{$Z\rightarrow l_1^- l_2^+$ decay in the model III with the
inclusion of non-universal extra dimensions.}
The extension of the Higgs sector in the SM brings new
contributions to the BRs of the processes and makes it possible to
obtain the flavor changing neutral current (FCNC) at tree level,
which plays an important role in the existence of flavor violating
(FV) interactions. Therefore, the multi Higgs doublet models are
worthwhile to study. The 2HDM is one of the candidate for the
multi Higgs doublet models. In the model I and II versions of the
2HDM, the FCNC at tree level is forbidden, however, those type of
interactions are possible in the model III version of the 2HDM.
The lepton flavor violating (LFV) Z decay $Z\rightarrow l_1^-
l_2^+$ can be induced at least in the one loop level in the
framework of the model III.

The addition of possible NUED, which are experienced by the gauge
bosons and the new Higgs particles, brings new contributions to
the BRs of the decays under consideration. In the model III, the
part of lagrangian which carries the interaction, responsible for
the LFV processes in 5 (6) dimension, reads
\begin{eqnarray}
{\cal{L}}_{Y}=
\xi^{D}_{5 (6)\, ij} \bar{l}_{i} \, (\phi_{2}|_{y (z)=0})\, E_{j}
+ h.c. \,\,\, , \label{lagrangian}
\end{eqnarray}
where
the couplings $\xi^{D}_{5 (6)\, ij}$ are $5 (6)$-dimensional
dimensionful Yukawa couplings which induce the LFV interactions.
These couplings can be rescaled to the ones in 4-dimension as
$\xi^{D}_{5 (6)\, ij}=\sqrt{2 \pi R}\,(2 \pi R)\, \xi^{D}_{ij}$
with lepton family indices $i,j$ \footnote{In the following, we
replace $\xi^{D}$ with $\xi^{D}_{N}$ where "N" denotes the word
"neutral"}. Here, $\phi_{2}$ is the new scalar doublet, R is the
compactification radius, $l_{i}$ and $E_{j}$ are lepton doublets
and singlets, respectively. The scalar and lepton doublets are the
functions of $x^\mu$ and $y$ ($y$, $z$), where $y$ ($z$) is the
coordinate represents the $5 (6)$'th dimension. Here we assume
that the Higgs doublet lying in the 4 dimensional brane has a
non-zero vacuum expectation value to ensure the ordinary masses of
the gauge fields and the fermions, however, the second doublet,
which is accessible to the extra dimensions, has no vacuum
expectation value, namely, we choose the doublets $\phi_{1}$ and
$\phi_{2}$ and the their vacuum expectation values as
\begin{eqnarray}
\phi_{1}=\frac{1}{\sqrt{2}}\left[\left(\begin{array}{c c}
0\\v+H^{0}\end{array}\right)\; + \left(\begin{array}{c c} \sqrt{2}
\chi^{+}\\ i \chi^{0}\end{array}\right) \right]\, ;
\phi_{2}=\frac{1}{\sqrt{2}}\left(\begin{array}{c c} \sqrt{2}
H^{+}\\ H_1+i H_2 \end{array}\right) \,\, , \label{choice}
\end{eqnarray}
and
\begin{eqnarray}
<\phi_{1}>=\frac{1}{\sqrt{2}}\left(\begin{array}{c c}
0\\v\end{array}\right) \,  \, ; <\phi_{2}>=0 \,\, .\label{choice2}
\end{eqnarray}
This choice ensures that the mixing between neutral scalar Higgs
bosons is switched off and it would be possible to separate the
particle spectrum so that the SM particles are collected in the
first doublet and the new particles in the second one
\footnote{Here $H^1$ ($H^2$) is the well known mass eigenstate
$h^0$ ($A^0$).}. Here we consider the gauge and $CP$ invariant
Higgs potential in two extra dimensions
\begin{eqnarray}
V(\phi_1, \phi_2 )&=& \delta(y)\,\delta(z)\, c_1\, (\phi_1^+
\phi_1-v^2/2)^2+ c_2\, (\phi_2^+ \phi_2)^2 +\nonumber \\ &+&
\delta(y)\,\delta(z)\, \Bigg( c_3 [(\phi_1^+ \phi_1-v^2/2) \,
(\phi_2^+ \phi_2)] + c_4 [(\phi_1^+ \phi_1)
(\phi_2^+ \phi_2)\nonumber \\
&-& (\phi_1^+ \phi_2)(\phi_2^+ \phi_1)] + c_5 [Re(\phi_1^+
\phi_2)]^2 + c_{6} [Im(\phi_1^+ \phi_2)]^2\Bigg) +c_{7} \, ,
\label{potential}
\end{eqnarray}
with constants $c_i, \, i=1,...,7$.

Since, only the new Higgs field $\phi_{2}$ is accessible to extra
dimensions in the Higgs sector, there appear KK modes
$\phi_{2}^{(n,r)}$ of $\phi_{2}$ in two spatial extra dimensions
after the compactification of the external dimensions on an
orbifold $(S^1\times S^1)/Z_2$ ,
\begin{eqnarray}
\phi_{2}(x,y,z ) & = & \frac{1}{(2 \pi R)^{d/2}} \left\{
\phi_{2}^{(0,0)}(x) + 2^{d/2} \sum_{n,r}^{\infty}
\phi_{2}^{(n,r)}(x) \cos(ny/R+rz/R)\right\} \, ,
\label{SecHiggsField}
\end{eqnarray}
where $d=2$, the indices  $n$ and $r$ are positive integers
including zero, but both are not zero at the same time. Here,
$\phi_{2}^{(0,0)}(x)$ is the 4-dimensional Higgs doublet which
includes the charged Higgs boson $H^+$, the neutral CP even (odd)
Higgs bosons $h^0$ ($A^0$). The KK modes of the charged Higgs
boson (neutral CP even (odd) Higgs $h^0$ ($A^0$)) have the mass
$\sqrt{m_{H^\pm}^2+m_n^2+m_r^2}$ ($\sqrt{m_{h^0}^2+m_n^2+m_r^2}$
($\sqrt{m_{A^0}^2+m_n^2+m_r^2}$) ), where $m_n=n/R$ and $m_r=r/R$.
Furthermore, we assume that the compactification radius $R$ is the
same for both new dimensions. Notice that the expansion for a
single extra dimension can be obtained by setting $d=1$, taking
$z=0$, and dropping the summation over $r$. In addition to the new
Higgs doublet, also the gauge fields feel the extra dimensions,
%
%
however, there is no additional contribution coming from the KK
modes of Z boson in the process under consideration since the Z
boson does not enter into calculations as an internal line. The
$Z-h^0$ KK mode-$A^0$ KK mode vertex is the same as the
4-dimensional one after the integration over extra dimensions,
except a small correction of the coupling due to the gauge field
$0$-mode-KK mode mixing (see section 6 for details).

Now, we would like to present the general effective vertex for the
interaction of on-shell Z-boson with a fermionic current:
\begin{eqnarray}
\Gamma_{\mu}=\gamma_{\mu}(f_V-f_A\ \gamma_5)+ \frac{i}{m_W}\,(f_M+f_E\,
\gamma_5)\, \sigma_{\mu\,\nu}\, q^{\nu}
\label{vertex}
\end{eqnarray}
where $q$ is the momentum transfer, $q^2=(p-p')^2$, $f_V$ ($f_A$)
is vector (axial-vector) coupling, $f_M$ ($f_E$) magnetic
(electric) transitions of unlike fermions. Here $p$
($-p^{\prime}$) is the four momentum vector of lepton
(anti-lepton) (see Fig. \ref{fig1ver} for the necessary 1-loop
diagrams due to neutral Higgs particles). Since the LFV Z boson
decay exists at least in the loop level, the KK modes of neutral
Higgs particles $h^0$ and $A^0$ contribute to the self energy and
vertex diagrams as internal lines. The leptons live in the 4D
brane and therefore they do not have any  KK modes. Notice that,
in the case of non-universal extra dimension, the KK number needs
not to be conserved and there exist $lepton-lepton-h^0$ KK mode
($A^0$ KK mode) vertices which can involve two zero modes and one
KK mode.

The  vector (axial-vector) $f_V$ ($f_A$) couplings and the
magnetic (electric) transitions $f_M$ ($f_E$) including the
contributions coming from a single extra dimension can be obtained
as
\begin{eqnarray}
f_V&=&f_V^{(0)}+2 \sum_{n=1}^{\infty} f_V^{(n)}\nonumber \, , \\
f_A&=&f_A^{(0)}+2 \sum_{n=1}^{\infty} f_A^{(n)}\nonumber \, , \\
f_M&=&f_M^{(0)}+2 \sum_{n=1}^{\infty} f_M^{(n)}\nonumber \, , \\
f_E&=&f_E^{(0)}+2 \sum_{n=1}^{\infty} f_E^{(n)}\, ,
\label{fVAMEex}
\end{eqnarray}
where $f^{(0)}_{V,A,M,E}$ are the couplings in the 4-dimensions
and $f^{(n)}_{V,A,M,E}$ are the ones due to the KK modes of the
scalar bosons $S=h^0, A^0$. The KK mode contributions
$f_{V,A,M,E}^{(n)}$ can be easily obtained by replacing the mass
squares $m^2_{S}$  in $f^{(0)}_{V,A,M,E}$ by $m^2_{S}+m_n^2$, with
$m_n=n/R$ and the compactification radius R. We present the
explicit expressions for the couplings $f^{(0)}_{V,A,M,E}$ in the
appendix, by taking into account all the masses of internal
leptons and external lepton (anti-lepton).

If we consider two NUED, the couplings $f_{V,A,M,E}^{(n)}$
appearing in eq. (\ref{fVAMEex}) should be replaced by
$f_{V,A,M,E}^{(n,r)}$ and the summation would be done over
$n,r=0,1,2 ...$ except $n=r=0$. Here $f_{V,A,M,E}^{(n,r)}$ can be
obtained by replacing the mass squares $m^2_{S}$  in
$f^{(0)}_{V,A,M,E}$ by $m^2_{S}+m_n^2+m_r^2$, with $m_n=n/R, \,
m_r=r/R$. Furthermore, the number $2$ in front of the summations
in eq. (\ref{fVAMEex}) would be replaced by $4$.

Finally, the BR for $Z\rightarrow l_1^-\, l_2^+$ can be written in
terms of the couplings $f_V$, $f_A$, $f_M$ and $f_E$ as
\begin{eqnarray}
BR (Z\rightarrow l_1^-\,l_2^+)=\frac{1}{48\,\pi}\,
\frac{m_Z}{\Gamma_Z}\, \{|f_V|^2+|f_A|^2+\frac{1}{2\,cos^2\,\theta_W}
(|f_M|^2+|f_E|^2) \}
\label{BR1}
\end{eqnarray}
where $\alpha_W=\frac{g^2}{4\,\pi}$ and $\Gamma_Z$ is the total
decay width of Z boson. In our numerical analysis  we consider the
BR due to the production of sum of charged states, namely
\begin{eqnarray}
BR (Z\rightarrow l_1^{\pm}\,l_2^{\pm})= \frac{\Gamma(Z\rightarrow
(\bar{l}_1\,l_2+\bar{l}_2\,l_1)}{\Gamma_Z} \, .\label{BR2}
\end{eqnarray}
%
\section{Discussion}
The LFV Z decays $Z\rightarrow e^{\pm} \mu^{\pm}$, $Z\rightarrow
e^{\pm} \tau^{\pm}$ and $Z\rightarrow \mu^{\pm} \tau^{\pm}$
strongly depend on the Yukawa couplings $\bar{\xi}^D_{N,ij}
\footnote{The dimensionfull Yukawa couplings
$\bar{\xi}^{D}_{N,ij}$ are defined as
$\xi^{E}_{N,ij}=\sqrt{\frac{4\,G_F}{\sqrt{2}}}\,
\bar{\xi}^{D}_{N,ij}$.}, i,j=e, \mu, \tau$ in the model III
version of 2HDM and these couplings are free parameters which
should be restricted by using the present and forthcoming
experiments. At first, we assume that the couplings which contain
at least one $\tau$ index are dominant similar to the Cheng-Sher
scenario \cite{Sher} and, therefore, we consider only the internal
$\tau$ lepton case among others. Furthermore, we assume that the
Yukawa couplings $\bar{\xi}^{D}_{N,ij}$ are symmetric with respect
to the indices $i$ and $j$. As a result, we need the numerical
values for the couplings $\bar{\xi}^{D}_{N,\tau e}$,
$\bar{\xi}^{D}_{N,\tau \mu}$ and $\bar{\xi}^{D}_{N,\tau \tau}$.

The upper limit of $\bar{\xi}^{D}_{N,\tau \mu}$ is predicted as
$30\, GeV$ (see \cite{Iltananomuon} and references therein) by
using the experimental uncertainty, $10^{-9}$, in the measurement
of the muon anomalous magnetic moment and assuming that the new
physics effects can not exceed this uncertainty. Using this upper
limit and the experimental upper bound of BR of $\mu\rightarrow e
\gamma$ decay, BR $\leq 1.2\times 10^{-11}$, the coupling
$\bar{\xi}^{D}_{N,\tau e}$ can be restricted in the range,
$10^{-3}-10^{-2}\, GeV$ (see \cite{Iltan1}). For the Yukawa
coupling $\bar{\xi}^{D}_{N,\tau \tau}$, we have no explicit
restriction region and we use the numerical values which are
greater than $\bar{\xi}^{D}_{N,\tau \mu}$. Furthermore, the
addition of the extra dimensions bring new parameter, namely the
compactification radius $R$ which arises from the compactification
of the a single (double) extra dimension on the orbifold $S^1/Z_2$
(($S^1\times S^1)/Z_2$).

In the present work, we study  the prediction of the NUED on the
BR of the LFV processes $Z\rightarrow l_1^{\pm} l_2^{\pm}$, in the
framework of the type III 2HDM. We see that the contribution
coming from two extra dimensions are considerably large compared
to the one coming from a single extra dimension, due to the crowd
of neutral scalar Higgs boson KK modes.

Throughout our calculations we use the input values given in Table
(\ref{input}).
\begin{table}[h]
        \begin{center}
        \begin{tabular}{|l|l|}
        \hline
        \multicolumn{1}{|c|}{Parameter} &
                \multicolumn{1}{|c|}{Value}     \\
        \hline \hline
        $m_{\mu}$                   & $0.106$ (GeV) \\
        $m_{\tau}$                   & $1.78$ (GeV) \\
        $m_{W}$             & $80.26$ (GeV) \\
        $m_{Z}$             & $91.19$ (GeV) \\
        $G_F$             & $1.16637 10^{-5} (GeV^{-2})$  \\
        $\Gamma_Z$                  & $2.490\, (GeV)$  \\
        $sin\,\theta_W$               & $\sqrt{0.2325}$ \\
        \hline
        \end{tabular}
        \end{center}
\caption{The values of the input parameters used in the numerical
          calculations.}
\label{input}
\end{table}

Fig. \ref{BrZemuksi} is devoted to  $\bar{\xi}^{D}_{N,\tau e}$
dependence of the  BR $\,(Z\rightarrow \mu^{\pm}\, e^{\pm})$ for
$\bar{\xi}^{D}_{N,\tau \mu}=1\,GeV$, $m_{h^0}=100\, GeV$ and
$m_{A^0}=200\, GeV$. The solid-dashed-small dashed lines represent
the BR without extra dimension-including a single extra dimension
for $1/R=500\, GeV$-including two extra dimensions for $1/R=500\,
GeV$. It is observed that the BR is not sensitive to the extra
dimension effects for a single extra dimension. However, for two
NUED, there is a considerable enhancement, almost two orders, in
the BR compared to the one without extra dimensions, even for the
small values of the coupling $\bar{\xi}^{D}_{N,\tau \mu}$. This is
due to the crowd of neutral Higgs boson KK modes. This enhancement
can be observed also in Fig. \ref{BrZemuR} where the BR is plotted
with respect to the compactification scale $1/R$ for
$\bar{\xi}^{D}_{N,\tau e}=0.05\, GeV$, $\bar{\xi}^{D}_{N,\tau
\mu}=1\,GeV$, $m_{h^0}=100\, GeV$ and $m_{A^0}=200\, GeV$. In this
figure the solid-dashed-small dashed lines represent the BR
without extra dimension-including a single extra dimension-
including two extra dimensions. It is seen that in the case of two
extra dimensions the BR reaches almost twice of the one without
extra dimensions, for the values of the compactification scale,
$1/R=2000\, GeV$. This enhancement becomes negligible for the
larger values of the compactification scales, $1/R>5000\, GeV$.
The possible enhancement due to the effect of two NUED on the
theoretical value of the BR of the corresponding Z decay is
worthwhile to study.

In Fig. \ref{BrZetauksi}, we present $\bar{\xi}^{D}_{N,\tau \tau}$
dependence of the  $BR\,(Z\rightarrow \tau^{\pm}\, e^{\pm})$ for
$\bar{\xi}^{D}_{N,\tau e}=0.05\,GeV$, $m_{h^0}=100\, GeV$ and
$m_{A^0}=200\, GeV$. The solid-dashed-small dashed lines represent
the BR without extra dimension-including a single extra dimension
for $1/R=500\, GeV$-including two extra dimensions for $1/R=500\,
GeV$. Similar to the previous process, the BR is not sensitive to
the extra dimension effects for a single extra dimension and this
sensitivity increases for two NUED. The enhancement of the BR of
two NUED is more than two orders larger compared to the one
without extra dimensions. Fig. \ref{BrZetauR} is devoted to the
compactification scale $1/R$ dependence of the BR for
$\bar{\xi}^{D}_{N,\tau \tau}=10\, GeV$, $\bar{\xi}^{D}_{N,\tau
\mu}=1\,GeV$, $m_{h^0}=100\, GeV$ and $m_{A^0}=200\, GeV$. In this
figure the solid-dashed-small dashed lines represent the BR
without extra dimension-including a single extra dimension-
including two extra dimensions. The enhancement in the BR for the
intermediate values of the compactification scale, namely $1/R\sim
1000\, GeV$, is more than one order. Similar to the previous
decay, this enhancement becomes small for the larger values of the
compactification scales, $1/R>5000\, GeV$.

Finally, Fig. \ref{BrZmutauksi} (\ref{BrZmutauR}) is devoted to
the $\bar{\xi}^{D}_{N,\tau \tau}$ (the compactification scale
$1/R$) dependence of the BR of the decay $Z\rightarrow
\tau^{\pm}\, \mu^{\pm}$ for $\bar{\xi}^{D}_{N,\tau \mu}=1\,GeV$
($\bar{\xi}^{D}_{N,\tau \mu}=1\,GeV$, $\bar{\xi}^{D}_{N,\tau
\tau}=10\,GeV$), $m_{h^0}=100\, GeV$ and $m_{A^0}=200\, GeV$. In
Fig. \ref{BrZmutauksi} the solid-dashed-small dashed lines
represent the BR without extra dimension-including a single extra
dimension for $1/R=500\, GeV$-including two extra dimensions for
$1/R=500\, GeV$. In the case of two extra dimensions, even for
small Yukawa couplings, it is possible to reach the experimental
upper limit of the BR of the corresponding decay, since the
enhancement in the BR is two order larger compared to the case
without extra dimensions. In Fig. \ref{BrZmutauR}, the
solid-dashed-small dashed lines represent the BR without extra
dimension-including a single extra dimension-including two extra
dimensions. It is observed that, in the case of two extra
dimensions, the BR reaches almost twice of the one without extra
dimensions, even for intermediate values of the compactification
scale, $1/R=2000\, GeV$. For the larger values of the
compactification scales, $1/R>5000\, GeV$, there is no enhancement
in the BR of the present decay.

At this stage we would like to present our results briefly.
\begin{itemize}
\item With the inclusion of a single NUED, the enhancement in the
BR of the LFV Z decays is small for the intermediate values of the
compactification scale $1/R$.
\item  In the case of two NUED, even for the small values of the
Yukawa couplings, it is possible to reach the experimental upper
limits of the BRs of the LFV Z  decays, since the enhancement in
the BR is two order larger compared to the case without extra
dimensions for the intermediate values of the compactification
scale $1/R$. This enhancement is due to the crowd of the KK modes
and it is an interesting result which may ensure an important
information to test the existence of the NUED, and if it exists,
to decide its number and to predict the lower limit of the
compactification scale, with the help of more accurate
experimental results.
\end{itemize}
As a summary, the effect of two NUED on the BRs of LFV Z decays
$Z\rightarrow l_1^{\pm}\, l_2^{\pm}$  is strong and the more
accurate future experimental results of these decays will be
useful to test the possible signals coming from the extra
dimensions.
\section{Acknowledgement}
This work has been supported by the Turkish Academy of Sciences in
the framework of the Young Scientist Award Program.
(EOI-TUBA-GEBIP/2001-1-8)
%
%
\section{The explicit expressions appearing in the text}
Here we present the explicit expressions for $f^0_V$, $f^0_A$,
$f^0_M$ and $f^0_E$ \cite{EiltZl1l2} (see eq. (\ref{fVAMEex})):
\begin{eqnarray}
f^0_V&=& \frac{g}{64\,\pi^2\,cos\,\theta_W} \int_0^1\, dx \,
\frac{1}{m^2_{l_2^+}-m^2_{l_1^-}} \Bigg \{ c_V \,
(m_{l_2^+}+m_{l_1^-})
\nonumber \\
&\Bigg(& (-m_i \, \eta^+_i + m_{l_1^-} (-1+x)\, \eta_i^V)\, ln \,
\frac{L^{self}_ {1,\,h^0}}{\mu^2}+ (m_i \, \eta^+_i - m_{l_2^+}
(-1+x)\, \eta_i^V)\, ln \, \frac{L^{self}_{2,\, h^0}}{\mu^2}
\nonumber \\ &+& (m_i \, \eta^+_i + m_{l_1^-} (-1+x)\, \eta_i^V)\,
ln \, \frac{L^{self}_{1,\, A^0}}{\mu^2} - (m_i \, \eta^+_i +
m_{l_2^+} (-1+x) \,\eta_i^V)\, ln \, \frac{L^{self}_{2,\,
A^0}}{\mu^2} \Bigg) \nonumber \\ &+&
c_A \, (m_{l_2^+}-m_{l_1^-}) \nonumber \\
&\Bigg ( & (-m_i \, \eta^-_i + m_{l_1^-} (-1+x)\, \eta_i^A)\, ln
\, \frac{L^{self}_{1,\, h^0}}{\mu^2} + (m_i \, \eta^-_i +
m_{l_2^+} (-1+x)\, \eta_i^A)\, ln \, \frac{L^{self}_{2,\,
h^0}}{\mu^2} \nonumber \\ &+& (m_i \, \eta^-_i + m_{l_1^-}
(-1+x)\, \eta_i^A)\, ln \, \frac{L^{self}_{1,\, A^0}}{\mu^2} +
(-m_i \, \eta^-_i + m_{l_2^+} (-1+x)\, \eta_i^A)\, ln \,
\frac{L^{self}_{2,\, A^0}}{\mu^2} \Bigg) \Bigg \} \nonumber \\ &-&
\frac{g}{64\,\pi^2\,cos\,\theta_W} \int_0^1\,dx\, \int_0^{1-x} \,
dy \, \Bigg \{ m_i^2 \,(c_A\,
\eta_i^A-c_V\,\eta_i^V)\,(\frac{1}{L^{ver}_{A^0}}+
\frac{1}{L^{ver}_{h^0}}) \nonumber \\ &-& (1-x-y)\,m_i\, \Bigg(
c_A\,  (m_{l_2^+}-m_{l_1^-})\, \eta_i^- \,(\frac{1}{L^{ver}_{h^0}}
- \frac{1}{L^{ver}_{A^0}})+ c_V\, (m_{l_2^+}+m_{l_1^-})\, \eta_i^+
\, (\frac{1}{L^{ver}_{h^0}} + \frac{1}{L^{ver}_{A^0}}) \Bigg)
\nonumber \\ &-& (c_A\, \eta_i^A+c_V\,\eta_i^V) \Bigg (
-2+(q^2\,x\,y+m_{l_1^-}\,m_{l_2^+}\, (-1+x+y)^2)\,
(\frac{1}{L^{ver}_{h^0}} +
\frac{1}{L^{ver}_{A^0}})-ln\,\frac{L^{ver}_{h^0}}{\mu^2}\,
\frac{L^{ver}_{A^0}}{\mu^2} \Bigg ) \nonumber \\ &-&
(m_{l_2^+}+m_{l_1^-})\, (1-x-y)\, \Bigg (
\frac{\eta_i^A\,(x\,m_{l_1^-} +y\,m_{l_2^+})+m_i\,\eta_i^-}
{2\,L^{ver}_{A^0\,h^0}}+\frac{\eta_i^A\,(x\,m_{l_1^-}
+y\,m_{l_2^+})- m_i\,\eta_i^-}{2\,L^{ver}_{h^0\,A^0}} \Bigg )
\nonumber \\ &+& \frac{1}{2}\eta_i^A\,
ln\,\frac{L^{ver}_{A^0\,h^0}}{\mu^2}\,
\frac{L^{ver}_{h^0\,A^0}}{\mu^2}
\Bigg \}\,, \nonumber \\
f^0_A&=& \frac{-g}{64\,\pi^2\,cos\,\theta_W} \int_0^1\, dx \,
\frac{1}{m^2_{l_2^+}-m^2_{l_1^-}} \Bigg \{ c_V \,
(m_{l_2^+}-m_{l_1^-})
\nonumber \\
&\Bigg(& (m_i \, \eta^-_i + m_{l_1^-} (-1+x)\, \eta_i^A)\, ln \,
\frac{L^{self}_{1,\,A^0}}{\mu^2} + (-m_i \, \eta^-_i + m_{l_2^+}
(-1+x)\, \eta_i^A)\, ln \, \frac{L^{self}_ {2,\,A^0}}{\mu^2}
\nonumber \\ &+& (-m_i \, \eta^-_i + m_{l_1^-} (-1+x)\,
\eta_i^A)\, ln \, \frac{L^{self}_{1,\, h^0}}{\mu^2}+ (m_i \,
\eta^-_i + m_{l_2^+} (-1+x)\, \eta_i^A)\, ln \,
\frac{L^{self}_{2,\,h^0}}{\mu^2} \Bigg) \nonumber \\ &+&
c_A \, (m_{l_2^+}+m_{l_1^-}) \nonumber \\
&\Bigg(& (m_i \, \eta^+_i + m_{l_1^-} (-1+x)\, \eta_i^V)\, ln \,
\frac{L^{self}_{1,\, A^0}}{\mu^2}- (m_i \, \eta^+_i + m_{l_2^+}
(-1+x)\, \eta_i^V)\, ln \, \frac{L^{self}_{2,\,A^0}}{\mu^2}
\nonumber \\ &+& (-m_i \, \eta^+_i + m_{l_1^-} (-1+x)\,
\eta_i^V)\, ln \, \frac{L^{self}_{1,\, h^0}}{\mu^2} + (m_i \,
\eta^+_i - m_{l_2^+} (-1+x)\, \eta_i^V)\, \frac{ln \,
L^{self}_{2,\,h^0}}{\mu^2} \Bigg) \Bigg \} \nonumber \\ &+&
\frac{g}{64\,\pi^2\,cos\,\theta_W} \int_0^1\,dx\, \int_0^{1-x} \,
dy \, \Bigg \{ m_i^2 \,(c_V\,
\eta_i^A-c_A\,\eta_i^V)\,(\frac{1}{L^{ver}_{A^0}}+
\frac{1}{L^{ver}_{h^0}}) \nonumber \\ &-& m_i\, (1-x-y)\, \Bigg(
c_V\, (m_{l_2^+}-m_{l_1^-}) \,\eta_i^- + c_A\,
(m_{l_2^+}+m_{l_1^-})\, \eta_i^+ \Bigg) \,(\frac{1}
{L^{ver}_{h^0}} - \frac{1}{L^{ver}_{A^0}}) \nonumber \\ &+& (c_V\,
\eta_i^A+c_A\,\eta_i^V) \Bigg(-2+(q^2\,x\,y-m_{l_1^-}\,m_{l_2^+}\,
(-1+x+y)^2) (\frac{1}{L^{ver}_{h^0}}+\frac{1}{L^{ver}_{A^0}})-
ln\,\frac{L^{ver}_{h^0}}{\mu^2}\,\frac{L^{ver}_{A^0}}{\mu^2}
\Bigg) \nonumber \\ &-& (m_{l_2^+}-m_{l_1^-})\, (1-x-y)\, \Bigg(
\frac{\eta_i^V\,(x\,m_{l_1^-} -y\,m_{l_2^+})+m_i\,\eta_i^+}
{2\,L^{ver}_{A^0\,h^0}}+ \frac{\eta_i^V\,(x\,m_{l_1^-}
-y\,m_{l_2^+})-m_i\, \eta_i^+}{2\,L^{ver}_{h^0\,A^0}}
\Bigg)\nonumber \\
&-& \frac{1}{2} \eta_i^V\, ln\,\frac{L^{ver}_{A^0\,h^0}}{\mu^2}\,
\frac{L^{ver}_{h^0\,A^0}}{\mu^2}
\Bigg \} \nonumber \,,\\
f^0_M&=&-\frac{g\, m_W}{64\,\pi^2\,cos\,\theta_W} \int_0^1\,dx\,
\int_0^{1-x} \, dy \, \Bigg \{ \Bigg( (1-x-y)\,(c_V\,
\eta_i^V+c_A\,\eta_i^A)\, (x\,m_{l_1^-} +y\,m_{l_2^+}) \nonumber
\\ &+& \, m_i\,(c_A\, (x-y)\,\eta_i^-+c_V\,\eta_i^+\,(x+y))\Bigg )
\,\frac{1}{L^{ver}_{h^0}} \nonumber \\ &+& \Bigg( (1-x-y)\, (c_V\,
\eta_i^V+c_A\,\eta_i^A)\, (x\,m_{l_1^-} +y\,m_{l_2^+})
-m_i\,(c_A\, (x-y)\,\eta_i^-+c_V\,\eta_i^+\,(x+y))\Bigg )
\,\frac{1}{L^{ver}_{A^0}} \nonumber \\ &-& (1-x-y) \Bigg
(\frac{\eta_i^A\,(x\,m_{l_1^-} +y\,m_{l_2^+})}{2}\, \Big (
\frac{1}{L^{ver}_{A^0\,h^0}}+\frac{1}{L^{ver}_{h^0\,A^0}} \Big )
+\frac{m_i\,\eta_i^-} {2} \, \Big ( \frac{1}{L^{ver}_{h^0\,A^0}}-
\frac{1}{L^{ver}_{A^0\,h^0}} \Big ) \Bigg ) \Bigg \} \,,\nonumber \\
f^0_E&=&-\frac{g\, m_W}{64\,\pi^2\, cos\,\theta_W} \int_0^1\,dx\,
\int_0^{1-x} \, dy \, \Bigg \{ \Bigg( (1-x-y)\,\Big ( -(c_V\,
\eta_i^A+c_A\,\eta_i^V)\, (x\,m_{l_1^-} -y\, m_{l_2^+}) \Big)
\nonumber \\ &-& m_i\, (c_A\,
(x-y)\,\eta_i^++c_V\,\eta_i^-\,(x+y))\Bigg )\,
\frac{1}{L^{ver}_{h^0}} \nonumber \\ &+& \Bigg ( (1-x-y)\,\Big (
-(c_V\, \eta_i^A+c_A\,\eta_i^V)\, (x\,m_{l_1^-} - y\, m_{l_2^+})
\Big ) + m_i\,(c_A\, (x-y)\,\eta_i^++c_V\,\eta_i^-\,(x+y)) \Bigg )
\,\frac{1}{L^{ver}_{A^0}} \nonumber \\&+& (1-x-y)\, \Bigg (
\frac{\eta_i^V}{2}\,(m_{l_1^-}\,x-m_{l_2^+}\, y)\, \, \Big (
\frac{1}{L^{ver}_{A^0\,h^0}}+\frac{1}{L^{ver}_{h^0\,A^0}} \Big )
+\frac{m_i\,\eta_i^+}{2}\, \Big (
\frac{1}{L^{ver}_{A^0\,h^0}}-\frac{1}{L^{ver}_{h^0\,A^0}} \Big )
\Bigg ) \Bigg \}, \label{fVAME}
\end{eqnarray}
where
\begin{eqnarray}
L^{self}_{1,\,h^0}&=&m_{h^0}^2\,(1-x)+(m_i^2-m^2_{l_1^-}\,(1-x))\,x
\nonumber \, , \\
L^{self}_{1,\,A^0}&=&L^{self}_{1,\,h^0}(m_{h^0}\rightarrow
m_{A^0})
\nonumber \, , \\
L^{self}_{2,\,h^0}&=&L^{self}_{1,\,h^0}(m_{l_1^-}\rightarrow
m_{l_2^+})
\nonumber \, , \\
L^{self}_{2,\,A^0}&=&L^{self}_{1,\,A^0}(m_{l_1^-}\rightarrow
m_{l_2^+})
\nonumber \, , \\
L^{ver}_{h^0}&=&m_{h^0}^2\,(1-x-y)+m_i^2\,(x+y)-q^2\,x\,y
\nonumber \, , \\
L^{ver}_{h^0\,A^0}&=&m_{A^0}^2\,x+m_i^2\,(1-x-y)+(m_{h^0}^2-q^2\,x)\,y
\nonumber \, , \\
L^{ver}_{A^0}&=&L^{ver}_{h^0}(m_{h^0}\rightarrow m_{A^0})
\nonumber \, , \\
L^{ver}_{A^0\,h^0}&=&L^{ver}_{h^0\,A^0}(m_{h^0}\rightarrow
m_{A^0}) \, , \label{Lh0A0}
\end{eqnarray}
and
\begin{eqnarray}
\eta_i^V&=&\xi^{D}_{N,l_1i}\xi^{D\,*}_{N,il_2}+
\xi^{D\,*}_{N,il_1} \xi^{D}_{N,l_2 i} \nonumber \, , \\
\eta_i^A&=&\xi^{D}_{N,l_1i}\xi^{D\,*}_{N,il_2}-
\xi^{D\,*}_{N,il_1} \xi^{D}_{N,l_2 i} \nonumber \, , \\
\eta_i^+&=&\xi^{D\,*}_{N,il_1}\xi^{D\,*}_{N,il_2}+
\xi^{D}_{N,l_1i} \xi^{D}_{N,l_2 i} \nonumber \, , \\
\eta_i^-&=&\xi^{D\,*}_{N,il_1}\xi^{D\,*}_{N,il_2}-
\xi^{D}_{N,l_1i} \xi^{D}_{N,l_2 i}\, . \label{etaVA}
\end{eqnarray}
The parameters $c_V$ and $c_A$ are $c_A=-\frac{1}{4}$ and
$c_V=\frac{1}{4}-sin^2\,\theta_W$. In eq. (\ref{etaVA}) the flavor
changing couplings $\bar{\xi}^{D}_{N, l_ji}$ represent the
effective interaction between the internal lepton $i$,
($i=e,\mu,\tau$) and outgoing (incoming) $j=1\,(j=2)$ one. Here
the couplings $\bar{\xi}^{D}_{N, l_ji}$ are complex in general and
they can be parametrized as
\begin{eqnarray}
\xi^{D}_{N,i l_j}=|\xi^{D}_{N,i l_j}|\, e^{i \theta_{ij}} \,\, ,
\label{xi}
\end{eqnarray}
where $i,l_j$ denote the lepton flavors and $\theta_{ij}$ are CP
violating parameters which are the possible sources of the lepton
EDM. However, in the present work we take these couplings real.
\newpage
\section{Gauge boson mass matrix and gauge coupling} \label{affaa}
Here we study an abelian model in the case of a single extra
dimension (two extra dimensions) with  two Higgs fields, where one
of Higgs field, $\phi_1(x)$,  is localized at the $y=0$ ($y=z=0$)
boundary of the $S^1/Z_2$ ($(S^1\times S^1)/Z_2$) orbifold and the
other one $\phi_1(x,y)$ ($\phi_1(x,y,z)$) is accessible to the
extra dimension(dimensions). Furthermore, we choose that only the
first Higgs field has a non-zero vacuum expectation value, and,
including a single extra dimension, the Higgs fields read:
\begin{eqnarray}
\phi_1(x)&=& \frac{1}{\sqrt{2}}
\Big( v+ h_1(x)+ i \,\chi_1(x)\Big) \,  \nonumber \\
\phi_2(x)&=&\frac{1}{\sqrt{2}} \Big( h_2(x,y)+ i
\,\chi_2(x,y)\Big) \label{AbelianHiggsField} \, .
\end{eqnarray}
Our aim is to obtain the gauge boson mass matrix, which is not
diagonal in the case of the non-universal extra dimensions where
the the gauge sector and the Higgs field $\phi_2$ is accessible to
the extra dimensions but the first Higgs field does not. In the
case of a single extra dimension, the detailed analysis on this
issue has been done in \cite{MuckPilaftsisRuckl} and both Higgs
fields assumed to have vacuum expectation values in that work. We
will present the crucial steps of this work briefly and we repeat
the same analysis for two extra dimensions.

The part of Lagrangian which carries the gauge and Higgs sector in
a single extra dimension reads
\begin{eqnarray}
{\mathcal L}(x,y) &=& - \, \frac{1}{4} \, F^{M N} \, F_{M N} +
(D_M \, \phi_2)^*\,(D^M \,
\phi_2)  + \delta (y) \, (D_{\mu} \, \phi_1)^*\, (D^{\mu} \, \phi_1)\nonumber\\
&&-\, V(\phi_1,\phi_2) \, + \, {\mathcal L}_{GF}(x,y)
\label{LagrAbelian}\, ,
\end{eqnarray}
where $V$  is   the CP and gauge invariant Higgs  potential,
$D_M = \partial_{M}  + i e_5 A_{M} (x,y)$, ($M=\mu,5$) is the
covariant derivative in 5 dimension and ${\mathcal L}_{GF}(x,y)$
is the gauge fixing term.

Now, we will present the sources of the gauge boson mass matrix in
the lagrangian eq. (\ref{LagrAbelian}): \\ \\
\textbf{1.} $F^{5 \mu} \, F_{5 \mu}$ in the part $F^{M N} \, F_{M N}$ \,,\\ \\
\textbf{2.} $(A_{\mu}\,A^{\mu})$ in the part $(D_{\mu} \,
\phi_1)^*\, (D^{\mu} \, \phi_1)$ \,,\\ \\
where $F^{M N}=\partial^N\, A^N-\partial^M\, A^N$, and, after
compactification, the gauge fields $A^N$ read
\begin{eqnarray}
A_{\mu}(x,y) & = & \frac{1}{(2 \pi R)^{1/2}} \left\{
A^{(0)}_{\mu}(x) + 2^{1/2} \sum_{n=1}^{\infty} A_{\mu}^{(n)}(x)
\cos(ny/R)\right\} \, ,
\nonumber \\
A_{5}(x,y) & = & \frac{1}{(2 \pi R)^{1/2}} \left\{ 2^{1/2}
\sum_{n=1}^{\infty} A_{5}^{(n)}(x) \sin(ny/R)\right\} \, .
\label{OneGaugField}
\end{eqnarray}
Notice that the $(D_M \, \phi_2)^*\,(D^M \, \phi_2)$ term in the
lagrangian (see eq. (\ref{LagrAbelian})) does not produce any mass
term for the gauge field since the scalar field $\phi_2$ does not
have any vacuum expectation value. The integration over the extra
dimension $y$ results in the mixing of zero mode and KK mode gauge
fields and the gauge boson mass matrix is obtained as (see
\cite{MuckPilaftsisRuckl}):
\begin{equation}
M_{A}^2 \, = \, \left(
\begin{array}{cccc}
m^2 & \sqrt{2} \, m^2 & \sqrt{2} \, m^2 & \cdots \\
\sqrt{2} \, m^2 & 2 \, m^2 \, + \, (1/R)^2 & 2 \, m^2 & \cdots \\
\sqrt{2} \, m^2 & 2 \, m^2 & 2 \, m^2 \, + \, (2/R)^2 & \cdots \\
\vdots & \vdots & \vdots & \ddots
\end{array}
\right) \quad ,
\end{equation}
where  $m^2  = e^2  v^2$. Using the determinant equation
\begin{equation}
\det \Big(M_{A}^2 - \lambda\,I \Big) = \Big(\prod_{n=1}^{\infty} (
n^2/R^2 - \lambda )\Big)\, \Big( m^2 - \lambda- 2 \, \lambda \,
m^2 \sum^{\infty}_{n=1} \, \frac{1}{(n / R)^2  - \lambda}\Big) = 0
\label{detone}
\end{equation}
the eigenvalues of the matrix is obtained by solving the
transcendental equation
\begin{equation}
 m_{A^{(n)}} = \pi \, m^2 \, R
\, \cot (\pi \, m_{A^{(n)}} \, R) \label{transcone}\, ,
\end{equation}
and KK  mass eigenstates $\hat{A}^{(n)}_{\mu}$ are given by
\begin{equation}
 \hat{A}_{\mu}^{(n)} \ = \ \bigg( 1
+\pi^2 \, m^2 \, R^2 + \frac{m^2_{A^{(n)}}}{m^2}\bigg)^{- 1/2} \,
\sum^{\infty}_{j=0} \, \frac{2 \, m_{A^{(n)}} \, m}{m^2_{A^{(n)}}
- (j / R)^2} \, \bigg(\frac{1}{\sqrt{2}}\bigg)^{\delta_{j,0}} \,
A_{\mu}{(j)} \, \label{masseigenstatesone}.
\end{equation}
For the non-abelian case the gauge field mass spectrum is
analogous to the abelian one presented above and the
transcendental equation for $Z$ boson is
\begin{equation}
 m_{Z^{(n)}} = \pi \, m_Z^2 \, R
\, \cot (\pi \, m_{Z^{(n)}} \, R) \label{transconemZ}\, ,
\end{equation}
and the corresponding coupling reads
\begin{equation}
g_{Z^{(n)}} =\sqrt{2} \, g \, \bigg(1+ \frac{m_Z^2}{m^2_{Z^{(n)}}}
+ \ \frac{\pi^2\, R^2\,m_Z^4}{m^2_{Z^{(n)}}}\bigg)^{-1/2}
\label{couplingZ}\, .
\end{equation}
At this stage we try to make the same analysis for two extra
dimensions. The part of Lagrangian which carries the gauge and
Higgs sector in two extra dimensions is
\begin{eqnarray}
{\mathcal L}(x,y) &=& - \, \frac{1}{4} \, F^{M N} \, F_{M N} +
(D_M \, \phi_2)^*\,(D^M \, \phi_2)  + \delta (y) \,\delta (z) \,
(D_{\mu} \, \phi_1)^*\, (D^{\mu} \, \phi_1)\nonumber\\ &&-\,
V(\phi_1,\phi_2) \, + \, {\mathcal L}_{GF}(x,y,z)
\label{LagrAbelian2}\, ,
\end{eqnarray}
$D_M = \partial_{M}  + i e_6 A_{M} (x,y,z)$, ($M=\mu,5,6$) is the
covariant derivative in 6 dimension. In this case the sources of
the gauge boson mass matrix in
the lagrangian (eq. (\ref{LagrAbelian2})) are: \\ \\
\textbf{1.} $F^{5 \mu} \, F_{5 \mu}$ and $F^{6 \mu} \, F_{6 \mu}$
in the part $F^{M N} \, F_{M N}$\\ \\
\textbf{2.} $(A_{\mu}\,A^{\mu})$ in the part $(D_{\mu} \,
\phi_1)^*\, (D^{\mu} \, \phi_1)$,
and, after compactification, the gauge fields $A_N$ read
\begin{eqnarray}
A_{\mu}(x,y) & = & \frac{1}{(2 \pi R)} \left\{ A^{(0, 0)}_{\mu}(x)
+ 2 \sum_{n,r}^{\infty} A_{\mu}^{(n,r)}(x) \cos(ny/R+rz/R)\right\}
\, ,
\nonumber \\
A_{5\, (6)}(x,y) & = & \frac{1}{(2 \pi R)} \left\{ 2
\sum_{n,r}^{\infty} A_{5\, (6)}^{(n,r)}(x) \sin(ny/R+rz/R)\right\}
\, , \label{TwoGaugField}
\end{eqnarray}
The integration over the extra dimensions $y$ and $z$ results in
the mixing of zero mode and KK mode gauge fields similar to the
one extra dimension case and the gauge boson mass matrix is
obtained as:
\begin{equation}
M_{A}^2 \, = \, \left(
\begin{array}{cccccc}
m^2 & 2 \, m^2 & 2 \, m^2 & 2 \, m^2 &2 \, m^2 & \cdots \\
2 \, m^2 & 4 \, m^2 \, + \, 2/R^2 & 4 \, m^2 &4 \, m^2 &4 \, m^2 &  \cdots \\
2 \, m^2 & 4 \, m^2 & 4 \, m^2 \, + \, 2/R^2 &4  \, m^2 &4 \, m^2 &\cdots \\
2 \, m^2 & 4 \, m^2 &4 \, m^2 & 4 \, m^2 \, + \, 4/R^2 & 4 \, m^2 &\cdots \\
2 \, m^2 & 4 \, m^2 &4 \, m^2 & 4 \, m^2 &4 \, m^2 + \, 10/R^2 &\cdots \\
\vdots & \vdots & \vdots& \vdots & \hspace{3cm}\ddots
\end{array}
\right) \quad ,
\end{equation}
with $m^2  = e^2  v^2$. Here, the mass spectrum is richer compared
to a single extra dimension case. Now the determinant equation
reads
\begin{equation}
\det \Big(M_{A}^2 - \lambda\,I \Big) = \Big(\prod_{n,r}^{\infty}
(\frac{2\,(n^2+r^2)}{R^2} - \lambda )\,\Big)\, \Big( m^2 -
\lambda- 4 \, \lambda \, m^2 \sum^{\infty}_{n,r} \,
\frac{1}{\frac{2\,(n^2+r^2)}{R^2} - \lambda }\Big) = 0
\label{dettwo} \, ,
\end{equation}
where the indices $n$ and $r$ are all positive integers including
zero, but both is not zero at the same time, and the
transcendental equation to obtain the eigenvalues of the matrix is
\begin{eqnarray}
 m^4_{A^{(p)}} &=& -(\frac{2\,m}{R})^2\,+(\frac{2}{R^2}+10\, m^2)\,
 \,m^2_{A^{(p)}}\,
 \,-(m_{A^{(p)}}\,m^2\,\pi\,R')\,\Big(
 \frac{1}{R'^2}-m_{A^{(p)}}^2
 \Big)\,\, \cot\, (\pi \, m_{A^{(p)}} \, R') \nonumber
\\ &-& (m^2_{A^{(p)}}\,m^2)\,\Big( \frac{1}{R'^2}-m_{A^{(p)}}^2
 \Big)\,\, \sum^{\infty}_{r=1}\,
\frac{\pi\, R'}{\sqrt{\lambda_r}}\,\, \cot \Big(\pi \,
\sqrt{\lambda_r} \, R'\Big ) \label{transctwo}\, ,
\end{eqnarray}
where $\lambda_r=m_{A^{(p)}}^2-r^2/R'^2$, $R'=R/\sqrt{2}$ and $p$
is the positive integer. Similar to the one extra dimension, for
the non-abelian case, the gauge field mass spectrum is analogous
to the abelian one presented above and the transcendental equation
is obtained by replacing the mass $m_{A^{(n)}}$ in eq.
(\ref{transctwo}) by $m_{Z^{(n)}}$. For two extra dimensions there
appears a new gauge coupling due to the complicated mass mixing,
however, in our numerical calculations, we used the one obtained
in the single extra dimension case by expecting that the new
contributions do not affect the behaviors of the physical
parameters we study. Notice that this coupling enters in the
calculations only for the zero mode Z boson case, since there is
no diagram which includes the KK mode virtual Z bosons.
\section{The vertices appearing in the present work}
In this section we present the vertices appearing in our
calculations. For a single extra dimension the Z boson gauge
coupling is given in eq. (\ref{couplingZ}) and, in the present
work, this coupling reads
\begin{equation}
g_{Z} =\sqrt{2} \, g \, \bigg(1+ \frac{m_Z^2}{m^2_{Z^{(0)}}} + \
\frac{\pi^2\, R^2\,m_Z^4}{m^2_{Z^{(0)}}}\bigg)^{-1/2}
\label{couplingZ0}\, ,
\end{equation}
where $m_{Z^{(0)}}$ is obtained by solving the eq.
(\ref{transconemZ}) for $n=0$. Furthermore, since there is no
mixing between neutral scalar Higgs bosons $H^0$ and $h^0$ due to
the our choice (see the section 2), the tree level interaction
$Z^{\mu}-H^0-A^0$ does not exist. Here  $L$ and $R$ denote chiral
projections $L(R)=1/2(1\mp \gamma_5)$, the parameters $c_{L\,(R)}$
read, $c_L=-1/2+s_W^2$, $c_R=s_W^2$ and $c_W=\cos\theta_W$,
$s_W=\sin\theta_W$, where $\theta_W$ is the weak angle.
\\ \\ \\ \\
\begin{figure}
\begin{tabular}{p{6cm} p{5cm}} \vskip -7.9truein
\parbox[b]{16cm}{\epsffile{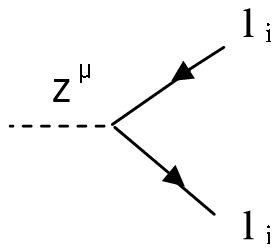}} &
\raisebox{5.ex}{$\frac{-i\,g_{Z}}{c_{W}}\,
\gamma^{\mu}\,\left[c_L\, L+c_R\, R\right]$}
\\ \\ \\ \vskip -7.0truein
\parbox[b]{6cm}{\epsffile{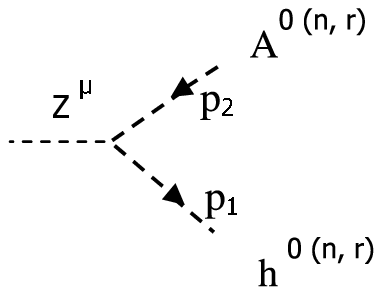}} & \\ \hspace{6.3cm}
\raisebox{5.ex} {$\frac{g_{Z}}{2\,c_{W}}\, (p_2+p_1)^{\mu}$}
\\ \\ \\ \vskip -7.2truein
\parbox[b]{6cm}{\epsffile{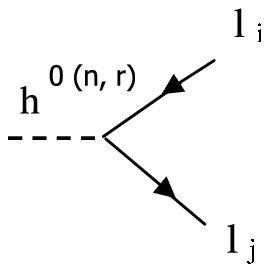}} & \vspace{0.7cm}\\ \hspace{6.3cm}
\raisebox{5.ex}{$\frac{-i}{2\sqrt{2}}\left[(\xi_{ij}^{D}+
\xi_{ji}^{D*})+(\xi_{ij}^{D}-\xi_{ji}^{D*})\gamma_5\right]$}\\
\\ \\ \vskip -7.2truein
\parbox[b]{6cm}{\epsffile{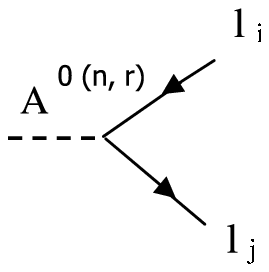}} & \vspace{0.7cm}\\ \hspace{6.3cm}
\raisebox{5.ex}{$\frac{1}{2\sqrt{2}}\left[(\xi_{ij}^{D}-
\xi_{ji}^{D*})+(\xi_{ij}^{D}+\xi_{ji}^{D*})\gamma_5\right]$}\\ \\
\\ \\ \\ \\
\end{tabular}
\caption{The vertices used in the present work.}
\end{figure}
%
%
%
\newpage
\newpage
\begin{figure}[htb]
\vskip 0.0truein \centering \epsfxsize=6.0in
\leavevmode\epsffile{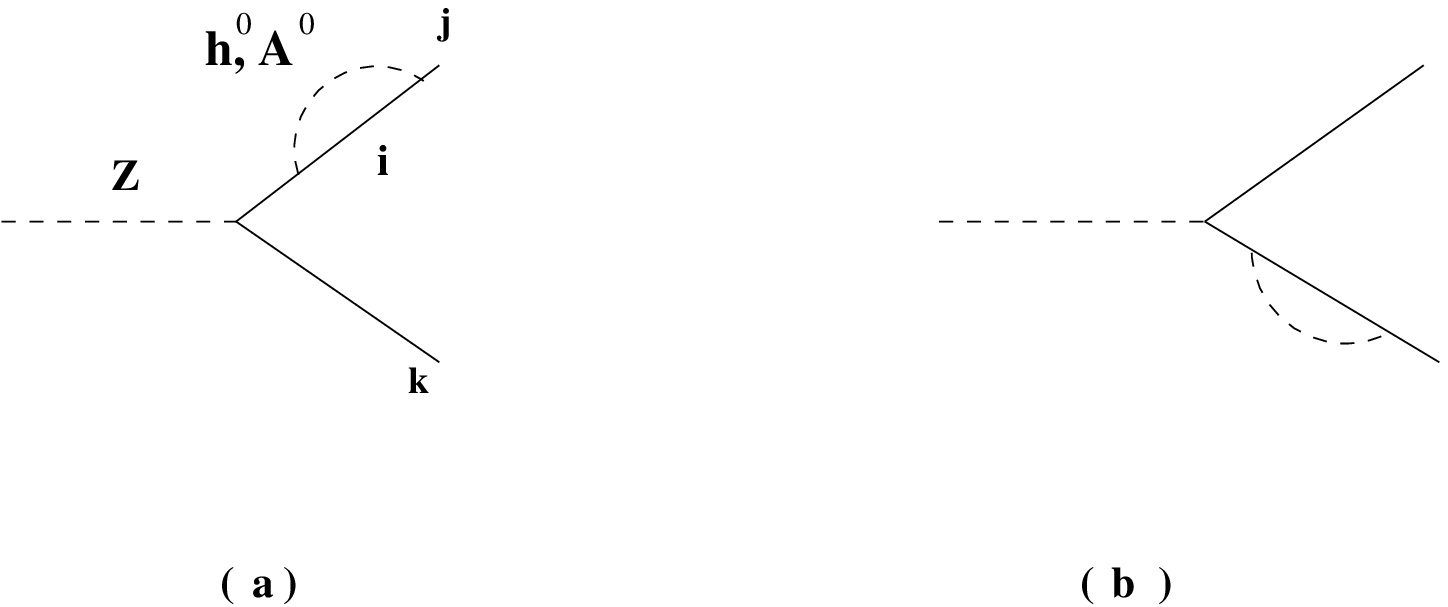} \vskip 0.5truein
\end{figure}
\begin{figure}[htb]
\vskip -0.5truein \centering \epsfxsize=6.0in
\leavevmode\epsffile{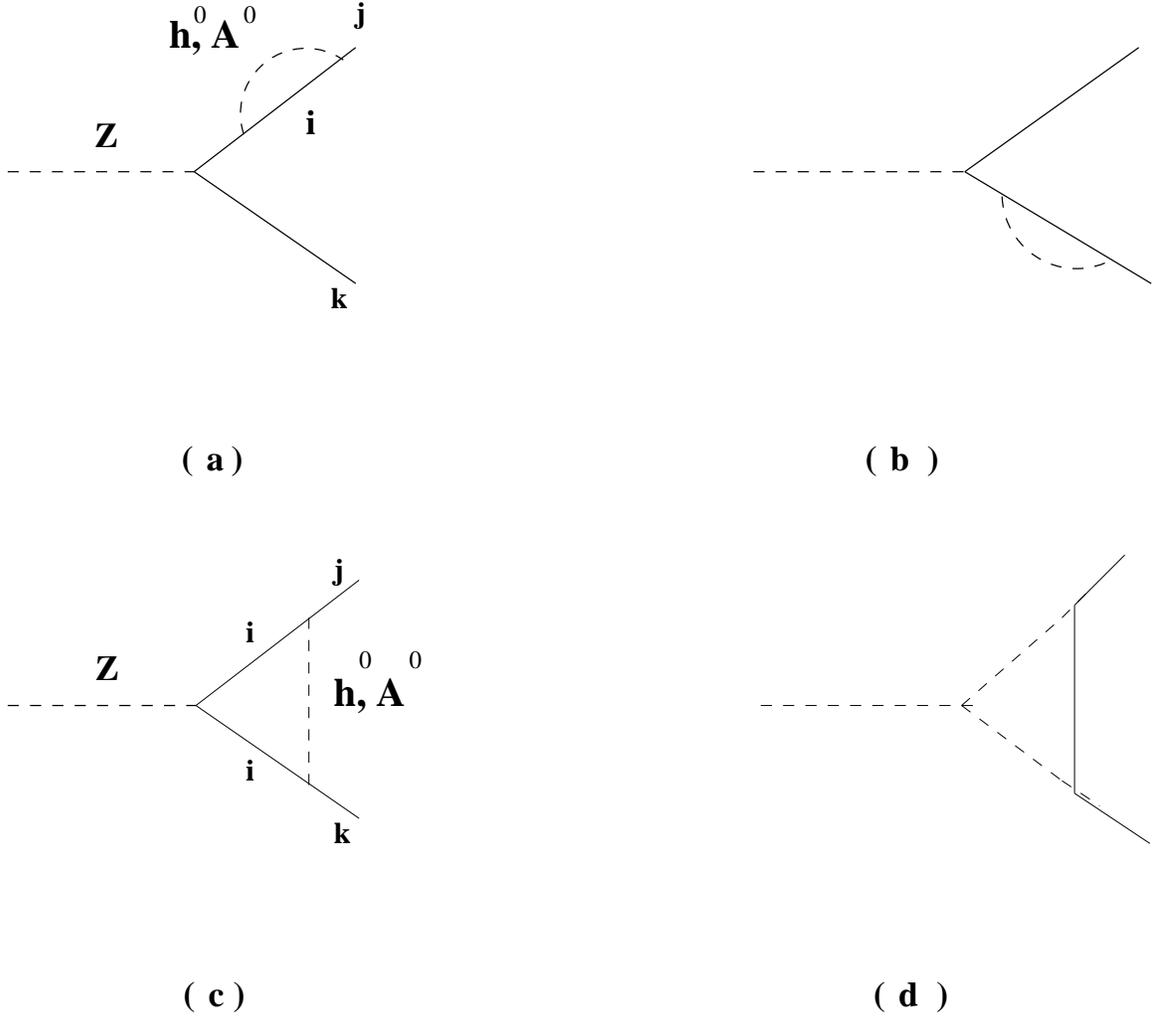} \vskip 0.5truein \caption[]{One
loop diagrams contribute to $Z\rightarrow k^+\,j^-$ decay due to
the neutral Higgs bosons $h_0$ and $A_0$ in the 2HDM. $i$
represents the internal, $j$ ($k$) outgoing (incoming) lepton,
dashed lines the vector field Z, $h_0$ and $A_0$ fields. In the
case 5 (6) dimensions the vertices are the same but there are
additional contributions due to the KK modes of $h_0$ and $A_0$
fields.} \label{fig1ver}
\end{figure}
\newpage
\begin{figure}[htb]
\vskip -3.0truein \centering \epsfxsize=6.8in
\leavevmode\epsffile{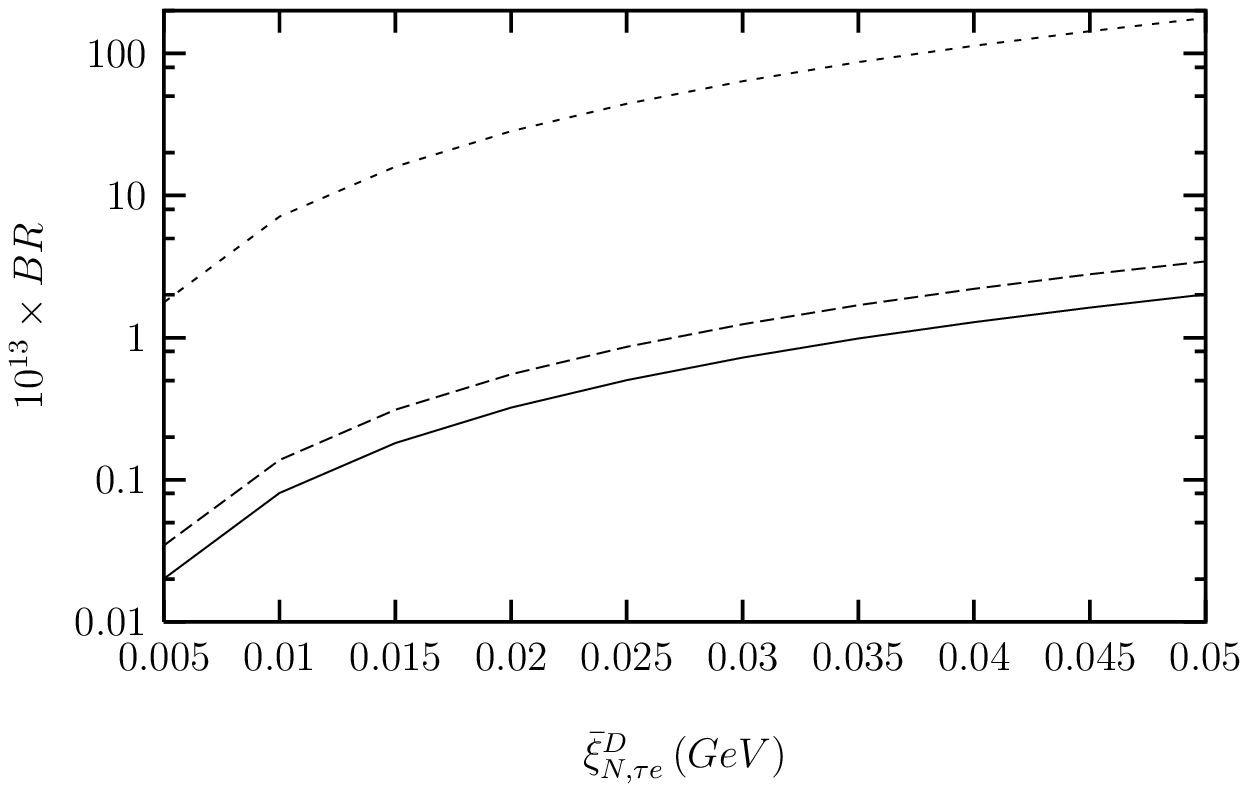} \vskip -3.0truein
\caption[]{$\bar{\xi}^{D}_{N,\tau e}$ dependence of the
$BR\,(Z\rightarrow \mu^{\pm}\, e^{\pm})$ for
$\bar{\xi}^{D}_{N,\tau \mu}=1\,GeV$, $m_{h^0}=100\, GeV$ and
$m_{A^0}=200\, GeV$. The solid-dashed-small dashed lines represent
the BR without extra dimension-including a single extra dimension
for $1/R=500\, GeV$-including two extra dimensions for $1/R=500\,
GeV$.} \label{BrZemuksi}
\end{figure}
\begin{figure}[htb]
\vskip -3.0truein \centering \epsfxsize=6.8in
\leavevmode\epsffile{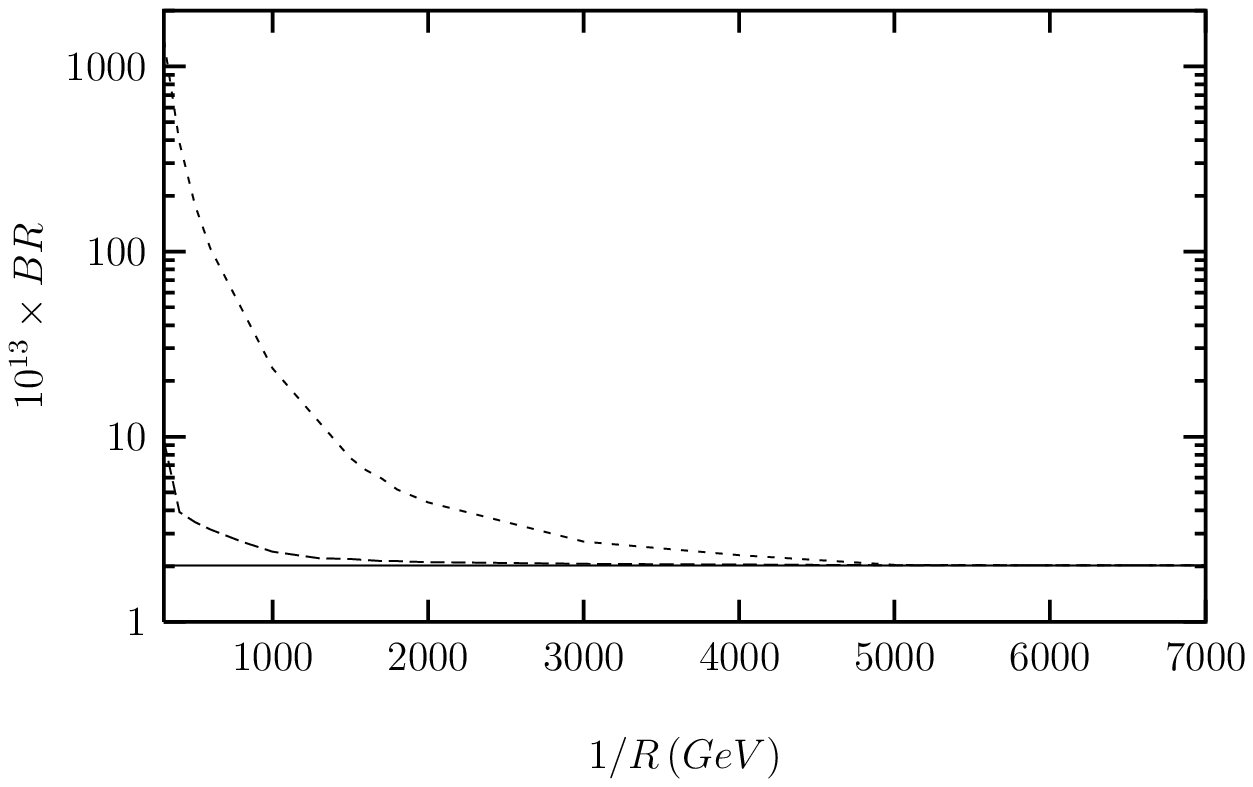} \vskip -3.0truein \caption[]{The
compactification scale $1/R$ dependence of $BR\,(Z\rightarrow
\mu^{\pm}\, e^{\pm})$ for $\bar{\xi}^{D}_{N,\tau e}=0.05\, GeV$,
$\bar{\xi}^{D}_{N,\tau \mu}=1\,GeV$, $m_{h^0}=100\, GeV$ and
$m_{A^0}=200\, GeV$. The solid-dashed-small dashed lines represent
the BR without extra dimension-including a single extra
dimension-including two extra dimensions.} \label{BrZemuR}
\end{figure}
\begin{figure}[htb]
\vskip -3.0truein \centering \epsfxsize=6.8in
\leavevmode\epsffile{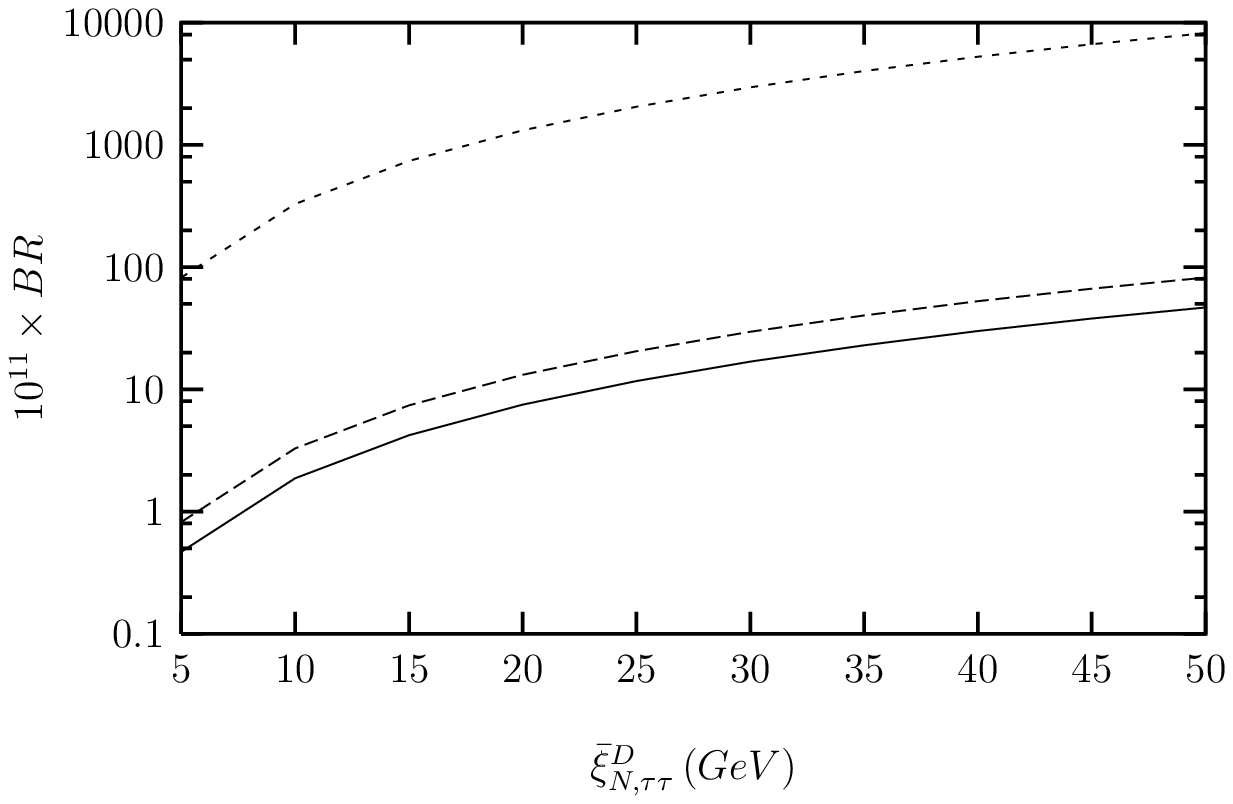} \vskip -3.0truein
\caption[]{$\bar{\xi}^{D}_{N,\tau \tau}$ dependence of the
$BR\,(Z\rightarrow \tau^{\pm}\, e^{\pm})$ for
$\bar{\xi}^{D}_{N,\tau e}=0.05\,GeV$, $m_{h^0}=100\, GeV$ and
$m_{A^0}=200\, GeV$. The solid-dashed-small dashed lines represent
the BR without extra dimension-including a single extra dimension
for $1/R=500\, GeV$-including two extra dimensions for $1/R=500\,
GeV$. } \label{BrZetauksi}
\end{figure}
\begin{figure}[htb]
\vskip -3.0truein \centering \epsfxsize=6.8in
\leavevmode\epsffile{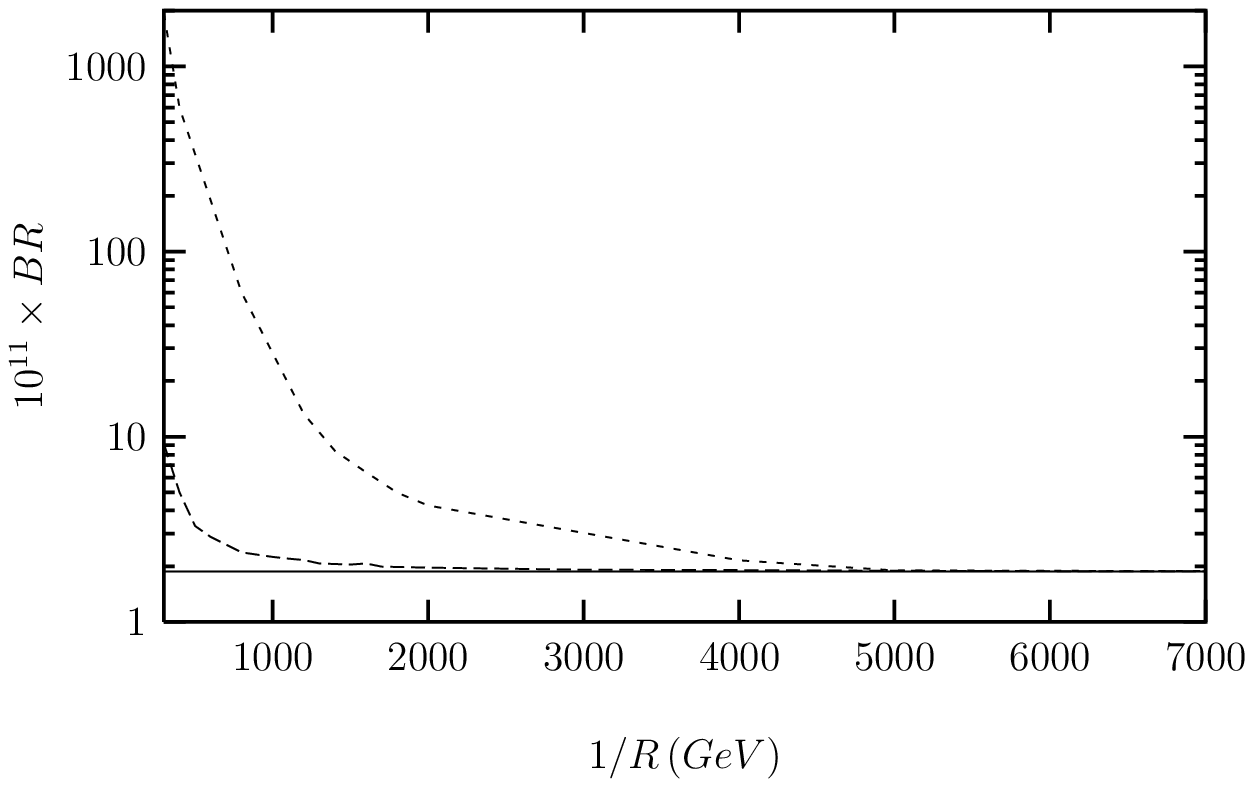} \vskip -3.0truein \caption[]{The
compactification scale $1/R$ dependence of the $BR\,(Z\rightarrow
\tau^{\pm}\, e^{\pm})$ for $\bar{\xi}^{D}_{N,\tau \tau}=10\, GeV$,
$\bar{\xi}^{D}_{N,\tau \mu}=1\,GeV$, $m_{h^0}=100\, GeV$ and
$m_{A^0}=200\, GeV$. The solid-dashed-small dashed lines represent
the BR without extra dimension-including a single extra
dimension-including two extra dimensions.} \label{BrZetauR}
\end{figure}
\begin{figure}[htb]
\vskip -3.0truein \centering \epsfxsize=6.8in
\leavevmode\epsffile{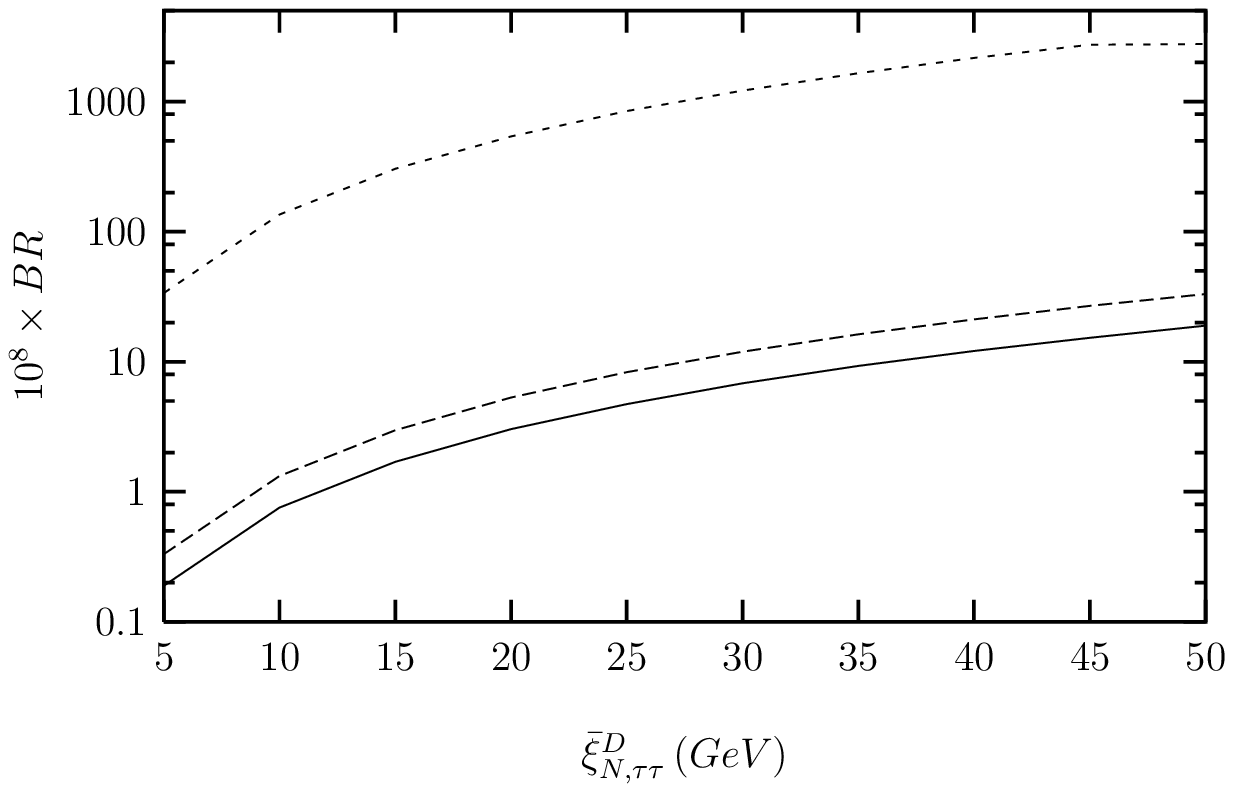} \vskip -3.0truein
\caption[]{The $\bar{\xi}^{D}_{N,\tau \tau}$ dependence of the BR
of the two decay $Z\rightarrow \tau^{\pm}\, \mu^{\pm}$ for
$\bar{\xi}^{D}_{N,\tau \mu}=1\,GeV$, $m_{h^0}=100\, GeV$ and
$m_{A^0}=200\, GeV$. The solid-dashed-small dashed lines represent
the BR without extra dimension-including a single extra dimension
for $1/R=500\, GeV$-including two extra dimensions for $1/R=500\,
GeV$.} \label{BrZmutauksi}
\end{figure}
\begin{figure}[htb]
\vskip -3.0truein \centering \epsfxsize=6.8in
\leavevmode\epsffile{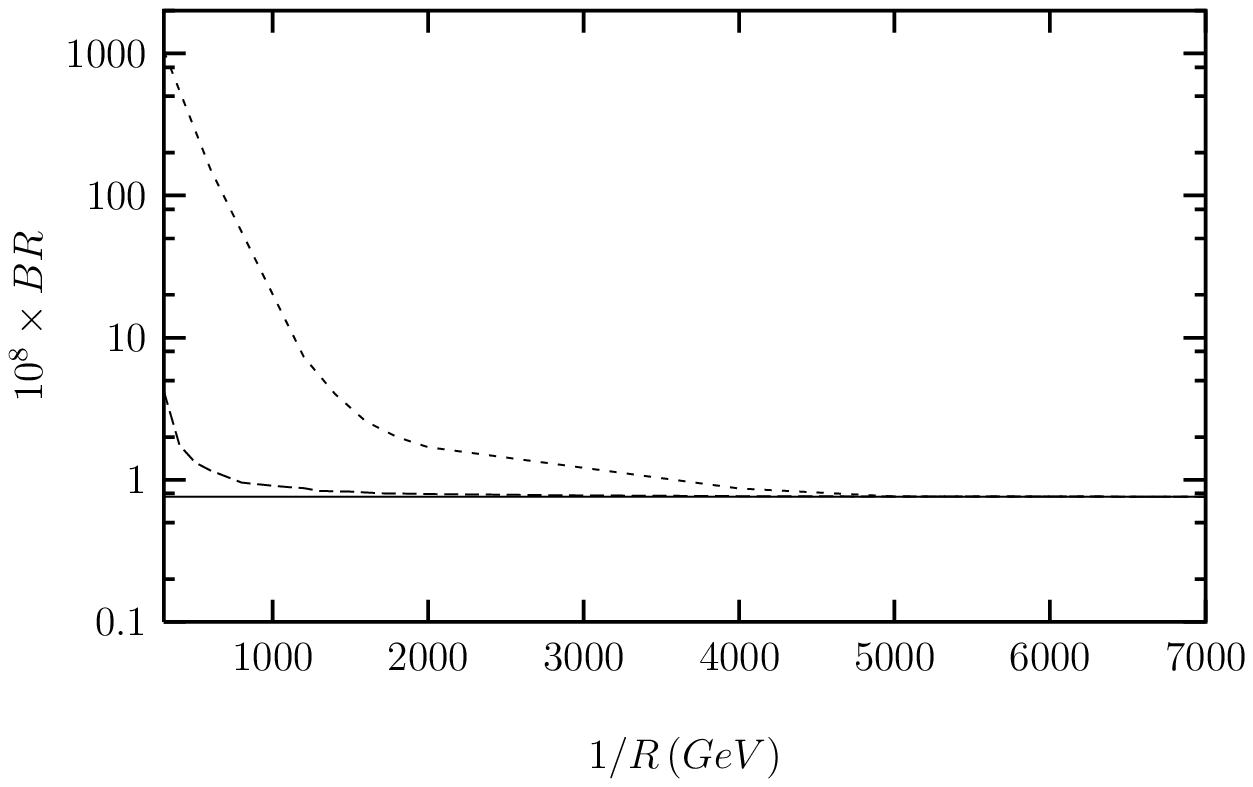} \vskip -3.0truein
\caption[]{The compactification scale $1/R$ dependence of the BR
of the two decay $Z\rightarrow \tau^{\pm}\, \mu^{\pm}$ for
$\bar{\xi}^{D}_{N,\tau \mu}=1\,GeV$, $\bar{\xi}^{D}_{N,\tau
\tau}=10\,GeV$, $m_{h^0}=100\, GeV$ and $m_{A^0}=200\, GeV$. The
solid-dashed-small dashed lines represent the BR without extra
dimension-including a single extra dimension-including two extra
dimensions. } \label{BrZmutauR}
\end{figure}
\end{document}